\documentclass[reprint, amsmath, amssymb, aps]{revtex4-1}	%revtex4-1
\usepackage{graphicx}
\usepackage{dcolumn}
\usepackage{bm}
\usepackage{mathrsfs}
\usepackage{tikz}
\usepackage{hyperref}
\input epsf.tex
\usepackage{amssymb}
\usepackage{amsmath}
\usepackage{amsthm}
\usepackage{amsopn}
\usepackage{ulem}
\usepackage{cancel}
\usepackage{color}
\usepackage{float}
\hypersetup{
    colorlinks=true,       % false: boxed links; true: colored links
    linkcolor=blue,          % color of internal links (change box color with linkbordercolor)
    citecolor=blue,        % color of links to bibliography
    filecolor=magenta,      % color of file links
    urlcolor=blue           % color of external links
}
\begin{document}

%\preprint{APS/123-QED}
\title{On detection statistics in double-double-slit experiment}

\author{M. J. Kazemi}
\email{kazemi.j.m@gmail.com}
\affiliation{Department of Physics, Faculty of Science, University of Qom , Qom, Iran}
\author{V. Hosseinzadeh}
\email{vahid.hoseinzade64@gmail.com}
\affiliation{School of Physics, Institute for Research in Fundamental Sciences (IPM), P.O. Box 19395-5531, Tehran, Iran}


\begin{abstract}
In this paper, we analyze the statistics of detection data in a general double-double-slit experiment. The two particles are detected at random times which are not equal in general and because we do not have any constraint on the distances of left and right screens from their slits and the ratio as well, they can be detected in completely different timescales. As the detection of first particle leads to collapse of the wave function, there is no a straightforward
and agreed method to study this problem in the orthodox
formalism which lacks a clear prediction of these random events and therefore the quantum state afterwards. 
This is not the case in Bohmian framework which we implement in this paper and we can predict the system up to the end of experiment. The main result is the joint distribution of detection data including the arrival time and position of the particles on left and right screens. As one of the main consequences, we see, although the joint spatial distribution can be affected by a change to the relative location of screens, the marginals on each side remain intact compatible with signal-locality. At the end, we see how this result is very sensitive to quantum equilibrium condition.

%The joint and marginal distributions of detection data are obtained in general 
	
%It is well-known that, in the orthodox interpretation of quantum mechanics, the prediction of the
%arrival/detection time distribution is ambiguous. In the entangled multi-particles systems, such as
%double-double-slit experiment, this ambiguity can leads to ambiguity on the position distribution.
%We discuss how the Bohmian mechanics can circumvent this obstacle and leads to a clear prediction.  Furthermore, despite the non-locality of Bohmian dynamics, we discussed how signal-locality is preserved.

\begin{description}
\item[PACS numbers]
03.65.Ta,  05.10.Gg
\end{description}

\end{abstract}
%\pacs{Valid PACS appear here}
\maketitle

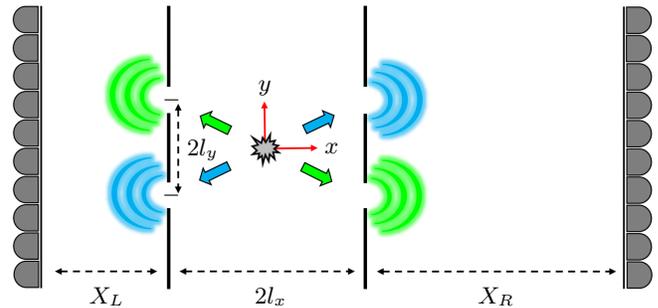
\begin{figure}[H]
	\centering
	\begin{tikzpicture}
		\draw (0,0) node[above right]{\includegraphics[width=1\linewidth, trim={0cm 0cm 0cm 0cm}]{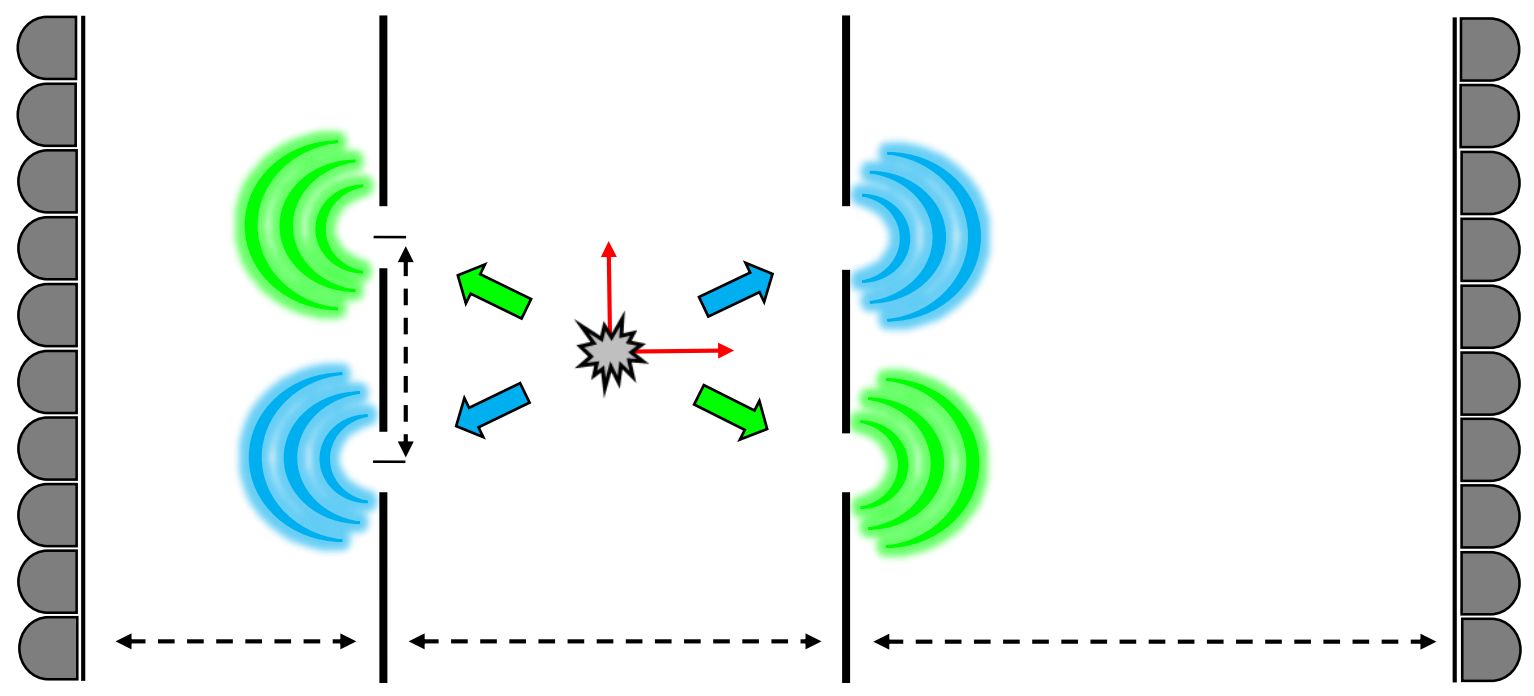}};
		\draw (3.3,-.15) node[above right]{$2l_x$};
		\draw (1.1,-.15) node[above right]{$X_L$};
		\draw (6.27,-.15) node[above right]{$X_R$};
		\draw (2.4,1.78) node[above right]{$2l_y$};
		\draw (4.23,1.88) node[above right]{$x$};
		\draw (3.35,2.65) node[above right]{$y$};
	\end{tikzpicture}
	\caption{\label{schematic} The source emits pairs of entangled particles, each of which passes through a double-slit. Two arrays of fast particle detectors are considered, on both sides, recording the detection events.}
	\label{DD-fig}
\end{figure}

%\printinunitsof{in}\prntlen{\textwidth}
%If the double-slit experiment be at the heart of the quantum theory \cite{}, the double-double-slit experiment must be at the head of it. \color
\textit{Introdution.---}Double-slit experiment has always been one of the central experiments in quantum mechanics. Yet, if there is an experiment more mysterious,  it is a double-double-slit experiment with  entangled particle pairs \cite{Horne1993}; see Fig(\ref{schematic}).  
%In foundations of quantum theory, if there is an experiment more mysterious than double-slit experiment,  it is double-double-slit experiment with  entangled particle pairs \cite{Horne1993}; see Fig(\ref{schematic}). 
In this experiment, the two most important properties of quantum theory, i.e. \textit{superposition} and \textit{entanglement}, combine and lead to a non-local interference pattern. The coincident measurements of two particles make interference fringes in the configuration space, while the detection of one of the individual particles does not show that \cite{Horne1989, Braverman2013}. This experiment has been realized experimentally using entangled photon pairs \cite{Kaur2020} or in another setup using entangled electron pairs on the molecular level \cite{Waitz2016}. It also can be carried out using pairs of metastable helium atoms as suggested in \cite{Kofler2012}.

%In general, however, the detection of the particles are not coincident (simultaneous)\cite{foot1}, and so the analysis of the experiment is more complex because of third quantum effect which is also present in the halfway of this  experiment, i.e. the \textit{collaps} of the wave function when one of the particle is detected by the screens.

%In this setup, in principle each particle is detected at different time and  one can experimentally measure two particle position-time joint probability distribution. %$P(\bm r_1,t_1;\bm r_2,t_2)$.
The full analysis of this experiment can be even more complex because of the third quantum effect, i.e. the \textit{collapse} of the wave function when the first particle is detected in the middle of the experiment at random times. This, in principle, can change the detection statistics of the remaining particle.  The analysis however is not completely straightforward in standard formalism. One of the reasons is that in the orthodox quantum mechanics we do not have a clear and agreed prediction of detection times distribution or the so-called arrival times of the first detected particle. 
%one needs to know when the collapse occurs. But, how one can compute the detection times distribution from initial wave 
%function? 
This problem is actually very old \cite{Allcock1969I,Kijowski1974,Werner1985}, but it is still open \cite{Vona2013,Das2021gauge,Das2021Questioning,Maccone2020,Anastopoulos2017}. The mathematical origin of this problem is that the time is a parameter in  the standard formalism, not a self-adjoint operator, hence, there is no unique and unambiguous way to compute the temporal probability distribution of detection events from first principles (i.e. Born rule) \cite{footPauli}.
% This in turn requires that we know the detection time and position of the first detected particle. 
% this question make a strong connection between  double-double-slit and the problem of arrival time in quantum theory.   
%This effect is often ignored in the theoretical analysis of this experiment while in a careful investigation,
  %This causes a problem as one can not predict clearly time and position of detection in orthodox quantum mechanics framework. 
%The effect of such collapse is introduced when the particle is measured at a given time, however,  in this situation, the detection time is random with undetermined distribution within orthodox quantum theory. 
Therefore, it is even not clear how one can determine the state of the system at a given time from initial two particle state;  because the state and its equation of motion change at an undetermined time. 
Although some aspects of this experiment have been explored, for example in \cite{Horne1989, Braverman2013, Kaur2020, Kofler2012, Waitz2016,Georgiev2021,Gneiting2013,Guay12003,Fonseca2000}, the analysis is not quite general and complete as they did not take the collapse effect into consideration and work in a regime where this effect has minimum and negligible consequences.

 %the previous theoretical study of this experiment this point is ignored, often due to the fact that the experiment is considered in a regime where this effect is small. Nonetheless, in general cases a deeper analysis would be needed.
%As far as we know, the previous theoretical study of this experiment is limited to eplaintion statistical results of coincident measurements,  where considering the collapse effect is not needed \cite{Horne1989, Braverman2013, Kaur2020, Kofler2012, Waitz2016,Georgiev2021,Gneiting2013,Guay12003,Fonseca2000}. Nonetheless, in general cases, a deeper analysis would be demanded.

% Therefore, even the position distribution at a specific time, which can be computed from the state via the Born rule, has a obscure status. 
 %Is the orthodox quantum mechanics a complete theory? 

%In this paper, we are going to study this problem via Bohmian mechanics which leads to a clear prediction in this situation. The non-relativistic Bohmian mechanics describes a world in which particles move along realistic trajectories.It has been proven that this theory is experimentally equivalent to the orthodox quantum mechanics \cite{} \textit{insofar as the latter is unambiguous} \cite{}; e.g., in usual position or momentum measurements at a specific time. In recent years, Bohmian mechanics has gained renewed interest for various reasons.One of this reasons is the fact that Bohmian trajectories can leads to the clear predictions for temporal distribution of detection  events \cite{}. In this regard, here we study double-double slit experiment via Bohmian framework. 

In this paper, we are going to study this experiment via Bohmian mechanics which leads to a clear prediction in this situation. In contrast to orthodox interpretation, the non-relativistic Bohmian mechanics describes a world in which particles move along realistic trajectories \cite{Bohm1952}. Nonetheless, it has been shown that this theory is experimentally equivalent to orthodox quantum mechanics \cite{Bohm1952,Durr2004} \textit{insofar as the latter is unambiguous} \cite{Bell1995, Ivanov2017, Das2019}; e.g., in usual position or momentum measurements at a specific time. Moreover, the Bohmian trajectories lead to a clear prediction for the temporal distribution of detection  events \cite{Das2019,Leavens1998,Zimmermann2016}. In fact, here we explain how one can compute two particles position-time joint probability distribution $P(\bm r_1,t_1;\bm r_2,t_2)$, the most general quantity that can be measured experimentally in this setup. Furthermore, we discuss how quantum statistics mask the non-locality of Bohmian dynamics, and signal-locality is preserved in this experiment.

 %In this regard, here we study double-double slit experiment via Bohmian framework. 

%In this paper, we are going to study this problem via Bohmian mechanics which leads to a clear prediction in this situation. The non-relativistic Bohmian mechanics describes a world in which particles move along realistic trajectories. This trajectories can be used to compute  temporal distribution of detection  events \cite{}.Note that, it has been proven that this theory is experimentally equivalent to the orthodox quantum mechanics \cite{} \textit{insofar as the latter is unambiguous} \cite{}; e.g., in usual position or momentum measurements at a specific time. In this regard, here we study double-double slit experiment via Bohmian framework.  

\textit{Theoretical framework and results.---}In the Bohmian mechanics, the state of a two-particle system is determined by the wave function $\Psi(\bm r_1,\bm r_2)$ and actual positions of particles $(\bm R_1,\bm R_2)$. 
%Before particle detection, 
The time evolution of the wave function, in free space, is given by the following two-particle Schrödinger equation
%\begin{equation}\label{Schrodinger}
%i\hbar\frac{\partial}{\partial t}\Psi_t=\sum_i\frac{-\hbar^2}{2m_i}\nabla^2_i\Psi_t%+V\Psi_t, 
%\end{equation}
\begin{equation}\label{Schrodinger}
i\hbar\frac{\partial}{\partial t}\Psi_t(\bm r_1,\bm r_2)=-(\frac{\hbar^2}{2m_1}\nabla^2_1+\frac{\hbar^2}{2m_2}\nabla^2_2)\Psi_t(\bm r_1,\bm r_2),
\end{equation}
where particle dynamics is given by the first-order differential equations in configuration space, the "guidance equation",  
\begin{align}\label{guiding}
\frac{d}{dt}\bm R_i(t)=\bm v_i^{\Psi_t}(\bm R_1(t),\bm R_2(t)),
\end{align} 
as $v_i^{\Psi_t}$ are the velocity fields  associated with the wave function $\Psi_t$; i.e. 
$v_i^{\Psi_t}{=}(\hbar/m_i)\Im(\nabla_i\Psi_t/\Psi_t)$ \cite{Quantum physics without quantum philosophy}.
When the particle $1$, for example, is detected at time $t=t_c$, the two-particle wave function 
collapses \textit{effectively} to a one-particle wave function as $\Psi_{t_c}(\bm r_1,\bm r_2)\to\psi_{t_c}(\bm r_2)$ where \cite{Durr2004,Durr2010On}:
\begin{equation}\label{collapsed wave function}
\psi_{t_c}(\bm r_2)=\Psi_{t_c}(\bm R_1(t_c),\bm r_2)
\end{equation}
known as “conditional wave function” in Bohmian formalism \cite{Quantum physics without quantum philosophy,Norsen2014}. 
For $t>t_c$, we can solve the one-particle Schrödinger equation
\begin{equation}\label{one particle Schrodinger}
i\hbar\frac{\partial}{\partial t}\psi_t(\bm r_2)=\frac{-\hbar^2}{2m_2}\nabla^2_2\psi_t(\bm r_2),
\end{equation}
with the initial wave function $\psi_{t_c}$ and also the associated one-particle guidance equations
\begin{equation}\label{one particle guiding}
\frac{d}{dt}\bm R_2(t)=\bm v_2^{\psi_t}(\bm R_2(t)),
\end{equation}
with the initial position $\bm R_2(t_c)$.

Here, in a double-double-slit setup, we restrict ourselves to propagation from the slits to detection screens. Hence, the initial wave function can be considered as follows \cite{Braverman2013,Georgiev2021}:
\begin{eqnarray}\label{Initial wave function}
\Psi_{t_0}(\bm r_1,\bm r_2)&=&[f_{u}^+(\bm r_1)f_{d}^-(\bm r_2)+f_{d}^+(\bm r_1)f_{u}^-(\bm r_2)] \nonumber\\
&+& 1\leftrightarrow 2,
\end{eqnarray}
where 
\begin{eqnarray}\label{fGG-def}
f_{u}^\pm(x,y)&=&G(x;\sigma_x,\pm l_x, \pm u_x)G(y;\sigma_y, +l_y,  +u_y),\nonumber\\
f_{d}^\pm(x,y)&=&G(x;\sigma_x,\pm l_x, \pm u_x)G(y;\sigma_y, -l_y,  -u_y),
\end{eqnarray}
and $G$ is a Gaussian wave function
$$G(x;\sigma, l,  u)=Ne^{-(x-l)^2/4 \sigma^2+ i m u (x-l)/\hbar}.$$
%in which the normalization factor is given by $N=[{(2 \pi)^{\frac14}\sqrt{\sigma_0}}]^{-1}$.
We also have symmetrized the wave function, as we have considered the particles as indistinguishable Bosons.  This kind of entangled Gaussian state is practical for implementation in quantum technologies because such states can be readily produced and reliably controlled \cite{Georgiev2021}. Furthermore, since the free two-particle Hamiltonian is separable, the time evolution of this wave function can be found from eq.\eqref{Schrodinger} analytically as
\begin{eqnarray}
    \Psi_{t}(\bm r_1,\bm r_2)&=&[f_{u}^+(\bm r_1,t)f_{d}^-(\bm r_2,t)+f_{d}^+(\bm r_1,t)f_{u}^-(\bm r_2,t)]\nonumber \\ 
&+& 1\leftrightarrow 2,
\end{eqnarray}
in which $f^{\pm}_{d/u}(t,\bm r)$ functions are constructed out of time dependent Gaussian wave functions $G_t$ as in \eqref{fGG-def} where
%where $f^{\pm}_{d/u}(t,\bm r)$ are defined in \eqref{fGG-def}, just initial Gaussian wave functions are replaced with its associated time dependent Gaussian wave functions,
%The time evolution of a single Gaussian wave function after solving the Schrodinger equation would be 
\begin{align}
    G_t(x;\sigma, l,  u) = (2 \pi s_t^2)^{-\frac14} e^{[\frac{-(x-l-u t)^2}{4\sigma s_t}]}e^{i\frac{m u}{\hbar}(x-l-\frac{ut}{2})}
\end{align}
and $s_t = \sigma (1+i  \frac{t\hbar}{2 m \sigma^2})$. 

In this paper, we run the simulation for $10^6$ trajectories in the configuration space and the parameters have been chosen as $\sigma_x=10^{-6}$(m), $\sigma_y=10^{-5}$(m), $u_x=0.1$(m/s),  $u_y=0$ (m/s), $l_x = 5\times10^{-3}$(m), $l_y = 5\times10^{-5}$(m)  and $m_1 = m_2 = 6.646 \times 10^{-27}$ (kg). These values are in agreement with the proposed setup in reference \cite{Kofler2012} in which colliding Helium-4 atoms have been used for producing initial entangled state  \cite{Perrin2007}.

%%%%%%%
\begin{figure}[H]
	\centering
	\begin{tikzpicture} 
		\draw (0,0) node[above right]{\includegraphics[width=1\linewidth, trim={0cm 0cm 0cm 0cm}]{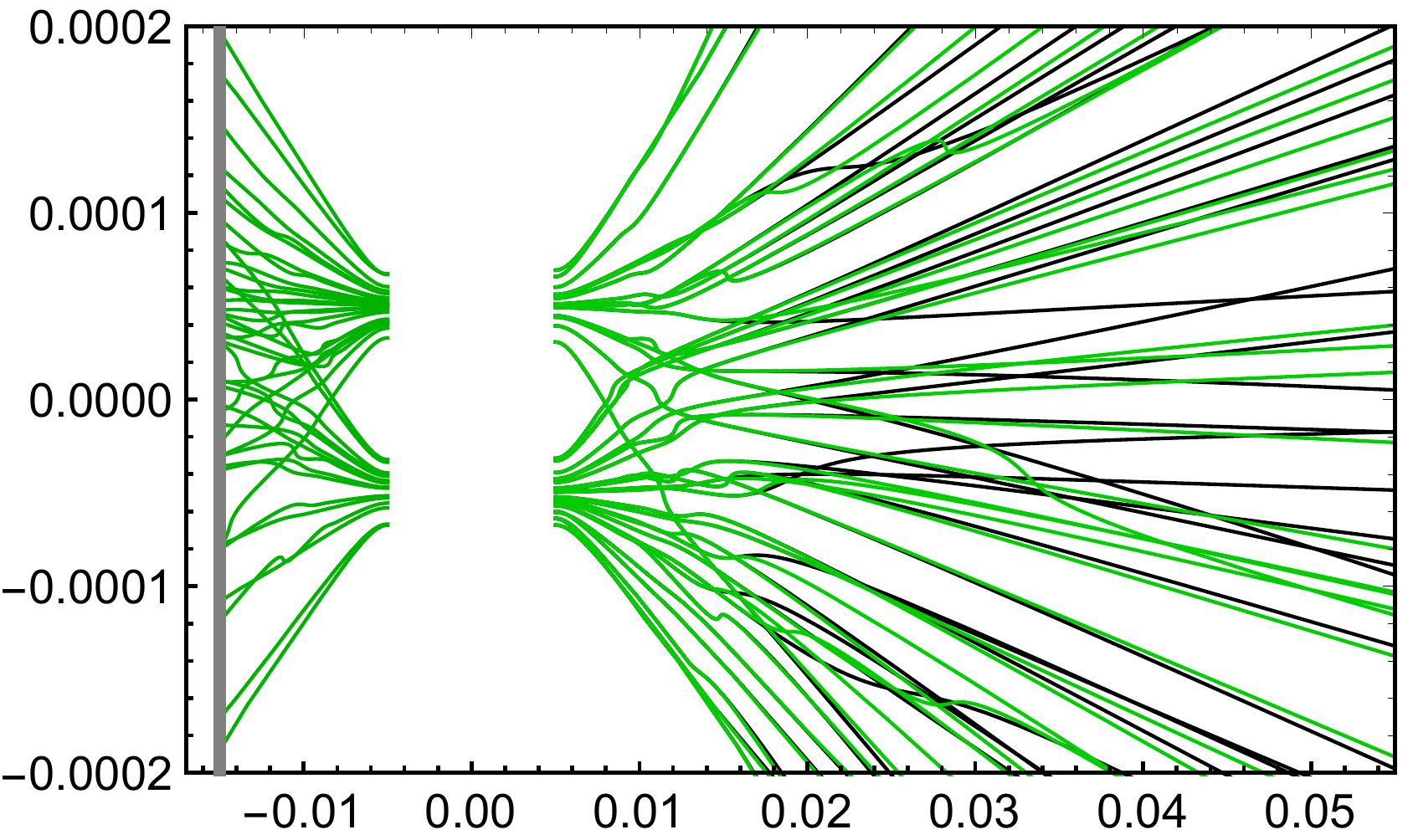}};
		 \draw (.3,5.3) node[above right]{$Y$(m)};
		 \draw (4.55,-.5) node[above right]{$X$(m)};
	\end{tikzpicture}
	\caption{The effect of collapse on trajectories. The green curves represent the case without considering collapse and the black ones are the modified trajectories due to the collapse effect. One can see that the right black trajectories are gradually deviate from green ones as the left screen detects the particle. Here, $X_R=0.5$(m) (which is beyond the box of this figure) and  $X_L=-0.015$(m) is represented by a vertical gray line.}\label{trajs-picture}
\end{figure}
%%%%%%%
The detection time and position of the first observed  particle are uniquely determined by solving equation \eqref{guiding}. Then, using equation \eqref{one particle guiding}, we can find the trajectories of the remaining particles. In figure \ref{trajs-picture}, we have depicted some of the trajectories for some random initial positions $\bm R^0 \equiv (\bm R_1(t_0),\bm R_2(t_0))$ sampled from $|\Psi_0(\bm R^0)|^2$ in accordance with the Born rule. In this figure the green trajectories are without considering the collapse effect and the black ones are with that. One can see that after the left particle detection, the right particle starts to deviate from the green paths as the conditional wave function now guides it. It is worth noting that, the ensemble of trajectories can be experimentally reconstructed using weak measurement techniques \cite{Braverman2013,Kocsis2011,Mahler2014,Mahler2016,Xiao2017,Traversa2013}, which can be used as a test for our result. 

\begin{figure*}[ht!]
\includegraphics[width=1.0\linewidth]{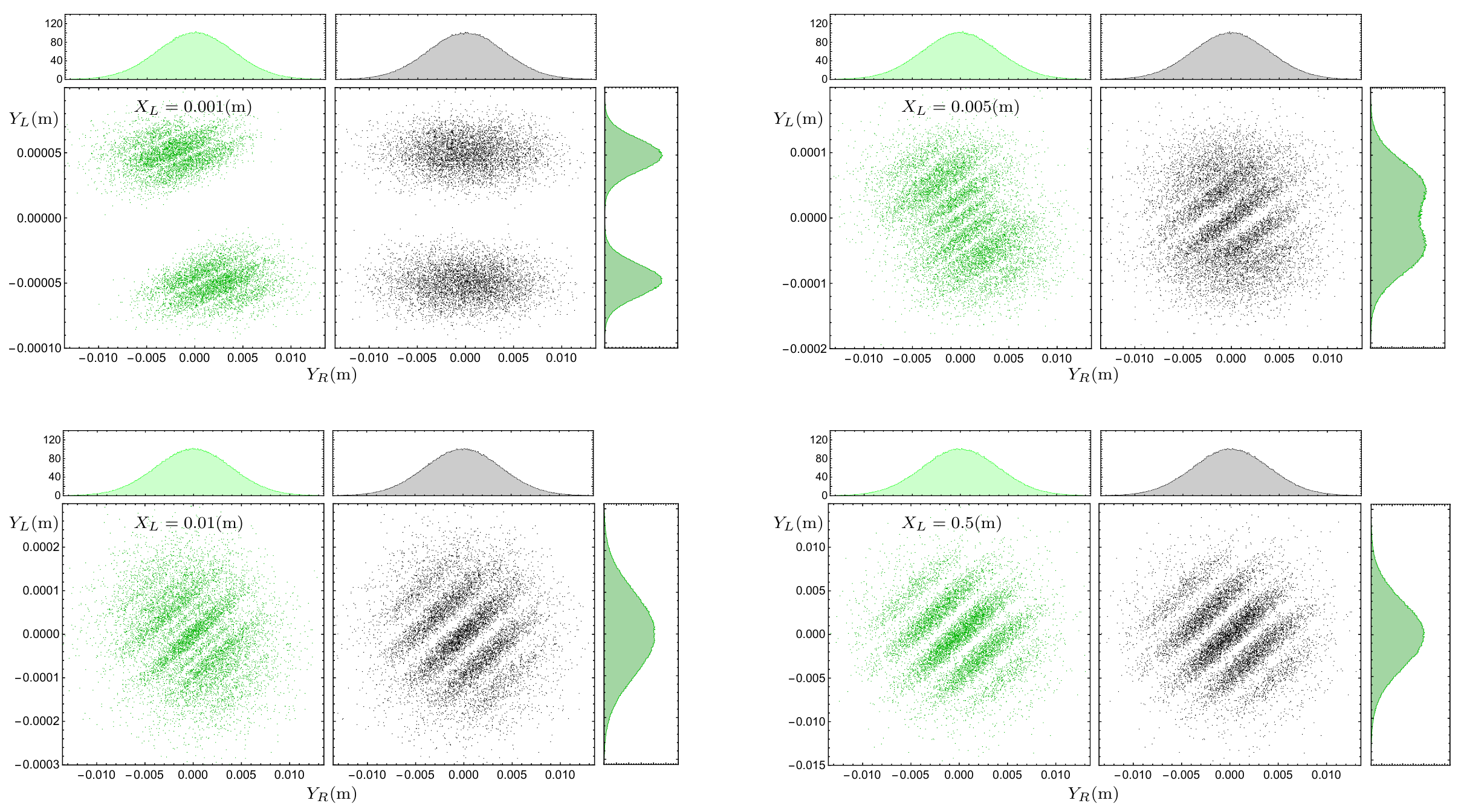}
\caption[font=\columnwidth, font=small]{
Here we have depicted joint distributions of $(y_L, y_R)$ data for different choices of $X_L$ where $X_R=0.5$ (m) for all the cases. Green ones are for the cases where we do not take the collapse into account and the black ones are the correct distributions with collapse. The right marginal and the left one are also depicted in the top and right of the joint distribution, respectively. The important point here is that the right marginal is independent of whether there is a measurement on the left or not and if there is a measurement, it is independent of the location of the left detection screen. Here, for the sake of better visualization, we have only included $10^4$ data points in the joint distributions. However, the marginal ones are computed with our $10^6$ data points. }\label{Joints-Marginals}
\end{figure*}

Using trajectories, we can find the joint detection data distribution in $(t_L,y_L;t_R,y_R)$ space where $t_{L/R}$ is the detection time $y_{L/R}$ is the detection position on the left/right screen.  The probability density behind this distribution can be formally written as
\begin{align*}
P(t_L,y_L;t_R,y_R)=\int d \bm R^0\ |\Psi_0(\bm R^0)|^2 \hspace{3cm}\\
\times \prod_{i=L,R}\delta(t_i-T_i(\bm R^0))\delta(y_i-Y_i{(\bm R^0)}),
\end{align*}
where  $T_{L,R}{(\bm  R^0_1, \bm  R^0_2)}$ and  $Y_{L,R}{(\bm  R^0_1, \bm  R^0_2)}$ are the arrival time and position of the particle with initial condition $(\bm  R^0_1, \bm  R^0_2)$ to the left and right screen, respectively. Note, how the above joint distribution and, therefore, any marginal distribution out of it, is sensitive to the Bohmian dynamics through functions $T$ and $Y$ and also to the Born rule by $|\Psi_0(\bm R^0)|^2$. We have plotted $(y_L,y_R)$ joint and also the right and left marginal data distributions in figure \ref{Joints-Marginals} for two cases: with collapse in dark and without that in green.

% After the first detection, the remaining system is in a mix state which can be formally represented by the following density matrix, 
%\begin{eqnarray*}
%	\hat\rho_t(\bm r_{1,2},\bm r'_{1,2}) = \sum_{\bm R^0} \psi_t(\bm r_{1,2}) \psi_t(\bm r'_{1,2})\,
%\end{eqnarray*}
% for one path of the above sum we see interference like double slit experiment. 
%Because from \eqref{collapsed wave function}, \eqref{one particle Schrodinger} and the fact that there is a negligible overlap between $f^{+}$ and $f^{-}$ wave packages, we can show the conditional wave function of the remaining system $\psi_t(\bm r_{1,2})$ as particle $2,(1)$ is detected first,
%\begin{align*}\label{Mix state}
%    \psi_t(\bm r_{1,2})\approx
%    \begin{cases}
%    A_+ f_{d}^-(\bm r_{1,2},t)+B_+ f_{u}^-(\bm r_{1,2},t) ,&  X_{2,1}(t_c) > 0\\
%    \\
%    A_- f_{d}^+(\bm r_{1,2},t)+B_- f_{u}^+(\bm r_{1,2},t) ,&  X_{2,1}(t_c) < 0
%    \end{cases}
%\end{align*}
%which is a superposition of the up and down slit states \footnote{Note that the exact wave function is sum of the two line in \eqref{Mix state} without any condition as we used in our code. However at each repetition one of the lines are empty waves having no role in the dynamics of particle.} like an ordinary double slit setup 
%where $A_\pm$ and $B_\pm$ are random constants determined by time and position of first particle detection $(t_c,\bm R(t_c))$,
%\begin{align*}
%    & A_{\pm} = f_{u}^\pm(\bm R_{2,1}(t_c),t_c), \,\,\,\, & B_{\pm} = f_{d}^\pm(\bm R_{2,1}(t_c),t_c) 
%\end{align*}
%Nonetheless , mixing paths washes out that pattern.

For more understanding of what is physically happening, note that from equations \eqref{collapsed wave function}, \eqref{one particle Schrodinger} and the fact that there is a negligible overlap between $f^{+}$ and $f^{-}$ wave packages, we can write the conditional wave function of the remaining system $\psi_t(\bm r_{1,2})$ as particle $2$, (particle $1$) is detected first,
\begin{align*}\label{Mix state}
    \psi_t(\bm r_{1,2})\approx
    \begin{cases}
    A_+ f_{d}^-(\bm r_{1,2},t)+B_+ f_{u}^-(\bm r_{1,2},t) ,&  X_{2,1}(t_c) > 0\\
    \\
    A_- f_{d}^+(\bm r_{1,2},t)+B_- f_{u}^+(\bm r_{1,2},t) ,&  X_{2,1}(t_c) < 0
    \end{cases}
\end{align*}
which is a superposition of the up and down slit states %\footnote{Note that the exact wave function is sum of the two line in \eqref{Mix state} without any condition as we used in our code. However at each repetition one of the lines are empty waves having no role in the dynamics of particle.} 
like an ordinary double-slit setup 
where $A_\pm$ and $B_\pm$ are random constants determined by time and position of the first particle detection $(t_c,\bm R(t_c))$,
\begin{align*}
   & A_{\pm} = f_{u}^\pm(\bm R_{2,1}(t_c),t_c), \,\,\,\, & B_{\pm} = f_{d}^\pm(\bm R_{2,1}(t_c),t_c) 
\end{align*}
As these numbers are random, the conditional wave function is random as well, and the system is really described by a density matrix which can be represented as,
\begin{eqnarray*}
	\hat\rho_t(\bm r_{1,2},\bm r'_{1,2}) = \sum_{\bm R^0} \psi_t(\bm r_{1,2}) \psi_t(\bm r'_{1,2}).
\end{eqnarray*}
So despite the fact that for one instance of this sum, quantum superposition leads to interference patterns like in the double-slit experiment, mixing them washes out that pattern, as we see in figure \ref{Joints-Marginals}. 
 
%\section{Signal-locality and quantum equilibrium}
%While quantum non-locality is manifest in Bohmian formalism, there is no room however for faster than light signaling. This can be rooted to the non-signaling theorem in standard quantum mechanics.
\begin{figure}[H]
	\centering
	\begin{tikzpicture}
		\draw (0,0) node[above right]{\includegraphics[width=.94\linewidth, trim={0cm 0cm 0cm 0cm}]{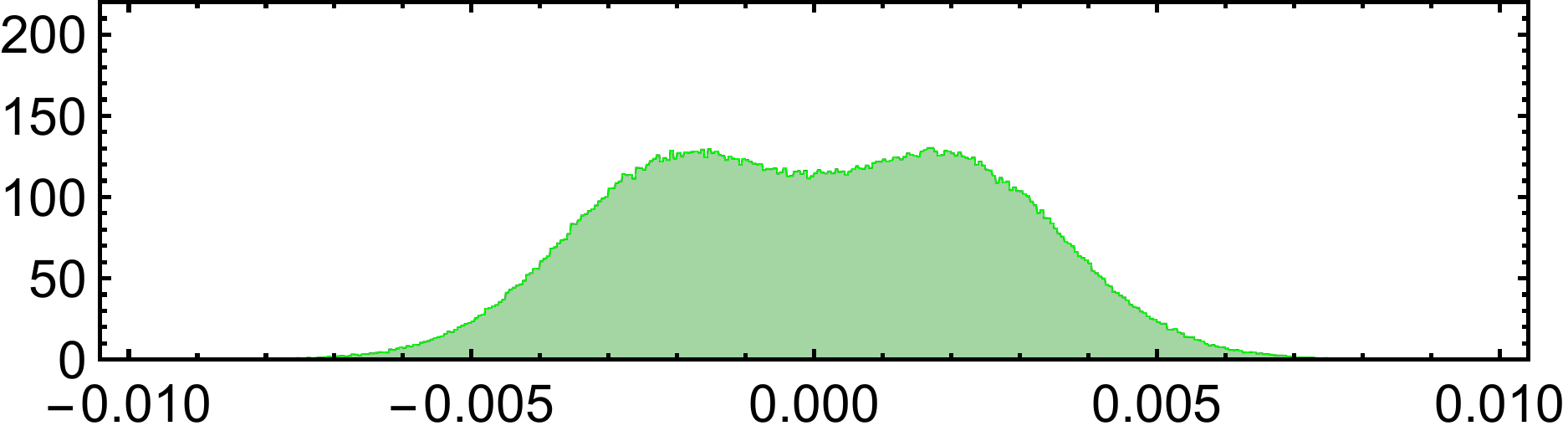}};
		\draw (0.8,1.7) node[above right]{$X_L=0.5$ (m)};
		%\draw (6.2,1.8) node[above right]{$X_R=0.5$ (m)};
		
		\draw (0,2.4) node[above right]{\includegraphics[width=.94\linewidth, trim={0cm 0cm 0cm 0cm}]{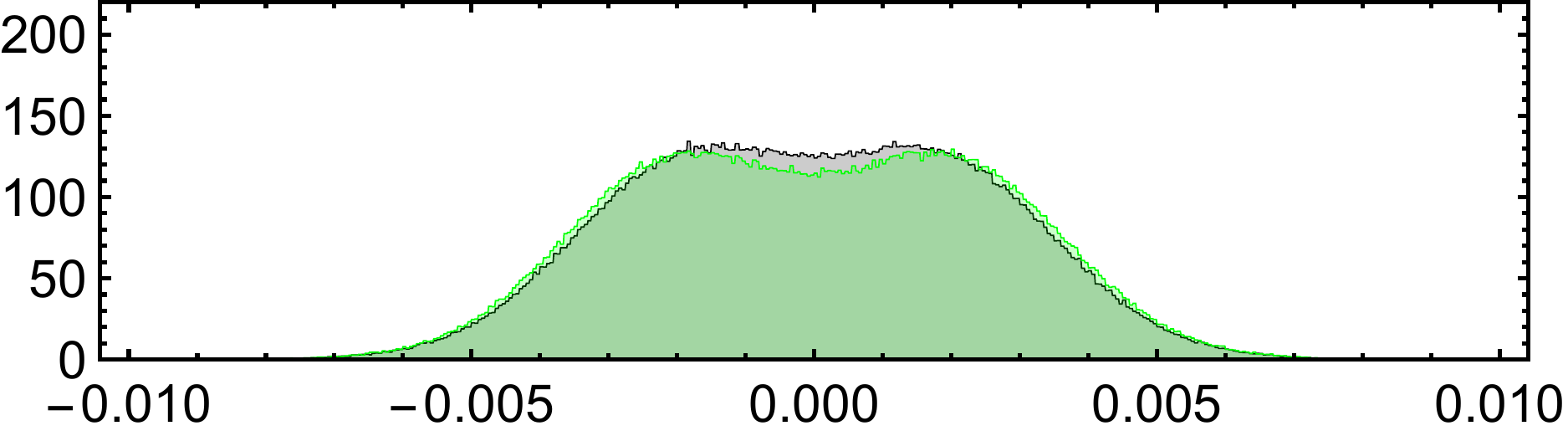}};
		\draw (0.8,4.1) node[above right]{$X_L=0.015$ (m)};
		%\draw (6.2,4.3) node[above right]{$X_R=0.5$ (m)};
		
		\draw (0,4.8) node[above right]{\includegraphics[width=.94\linewidth, trim={0cm 0cm 0cm 0cm}]{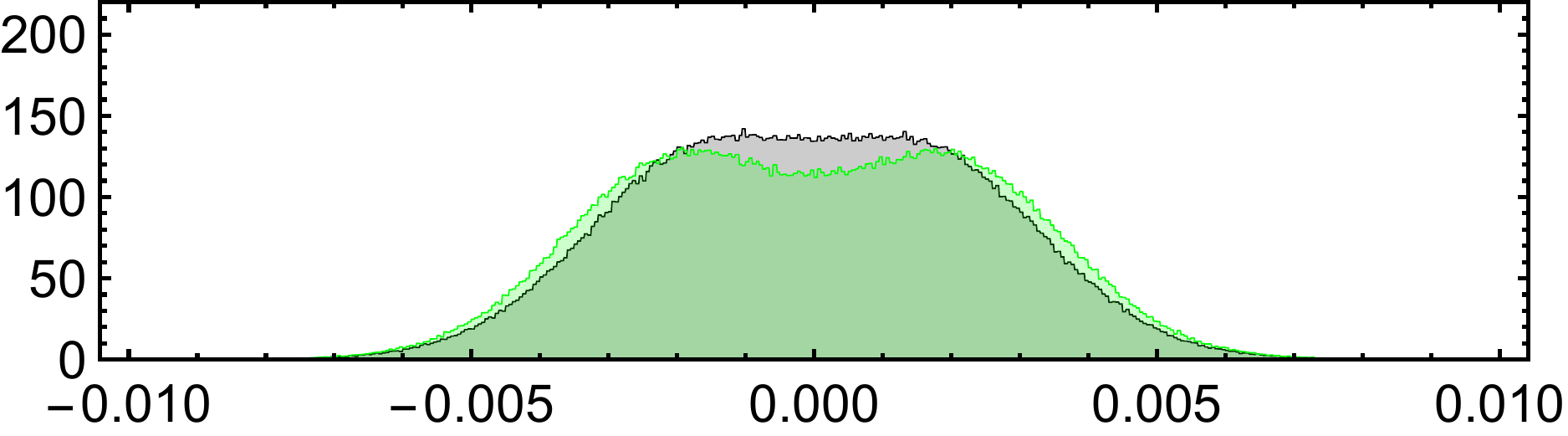}};
		\draw (0.8,6.5) node[above right]{$X_L=0.011$ (m)};
		%\draw (6.2,6.8) node[above right]{$X_R=0.5$ (m)};
		
		\draw (0,7.2) node[above right]{\includegraphics[width=.94\linewidth, trim={0cm 0cm 0cm 0cm}]{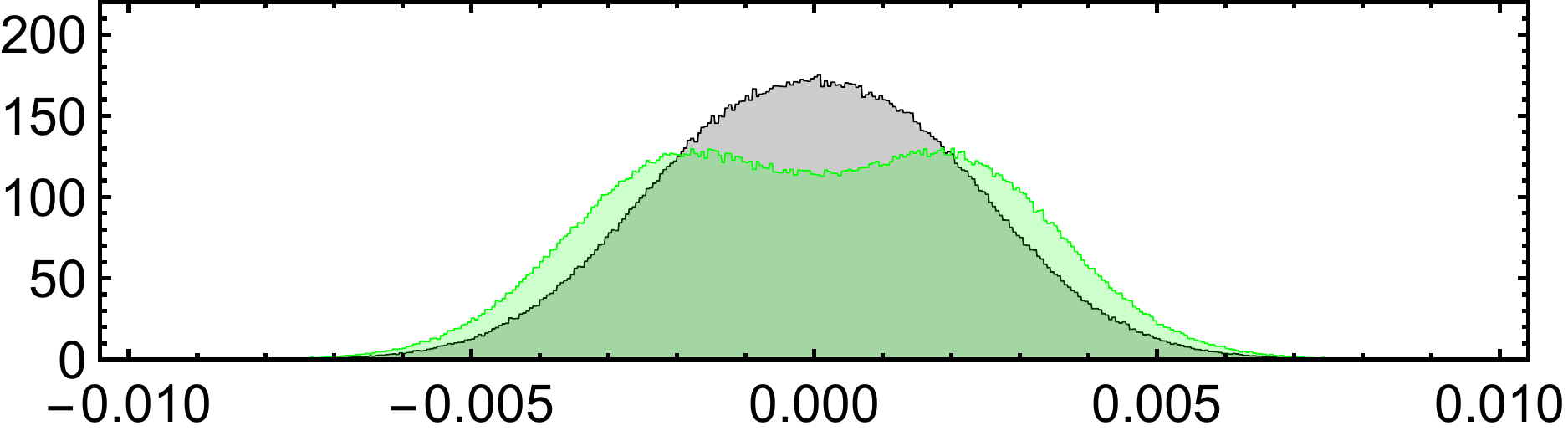}};
		\draw (0.8,8.9) node[above right]{$X_L=0.007$ (m)};
		%\draw (6.2,9.3) node[above right]{$X_R=0.5$ (m)};
		
		\draw (0,9.6) node[above right]{\includegraphics[width=.94\linewidth, trim={0cm 0cm 0cm 0cm}]{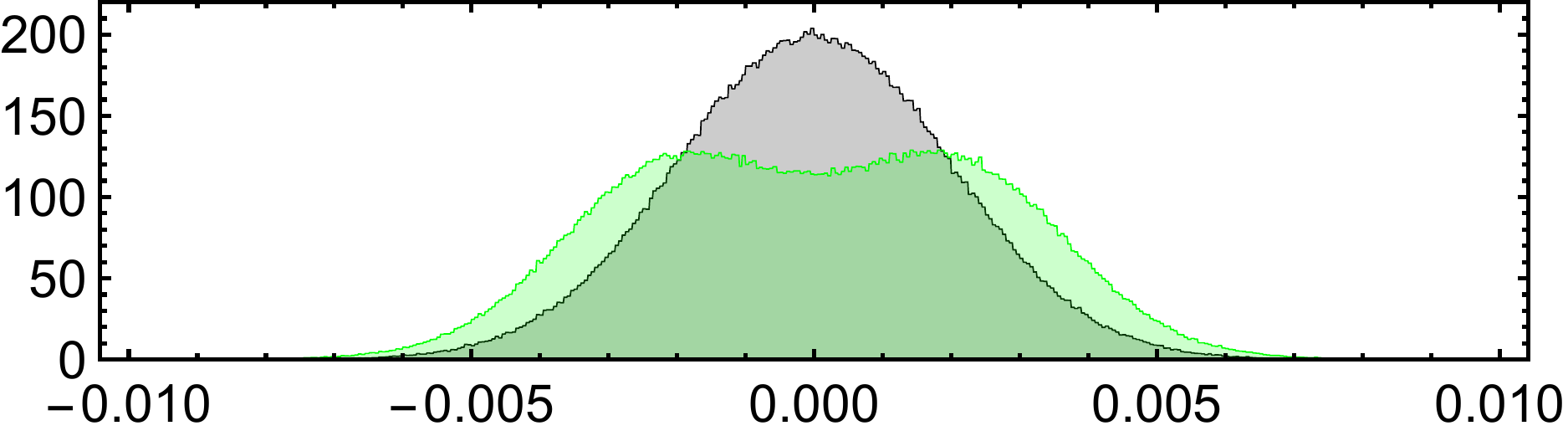}};
		\draw (0.8,11.3) node[above right]{$X_L=0.001$ (m)};
		%\draw (6.2,11.8) node[above right]{$X_R=0.5$ (m)};
		%%%%%%%%%%
		\draw (3.83,-.5) node[above right]{$Y$(m)};
		\draw (.05,4) node[above right, rotate=90]{{Arrival Position distribution}};
	\end{tikzpicture}
	\caption{Distribution of arrival position on the right screen. The different panels represent the arrival position distributions for different values of $X_L$. In all cases, the right screen surface is located at $X_R=0.5$(m). As we expect the dark and green distribution coincide when $X_L=X_R$ as there is not much for the remaining particle to be guided by the conditional wave function.} 
	\label{Deviation from Born Interference-fig}
\end{figure}
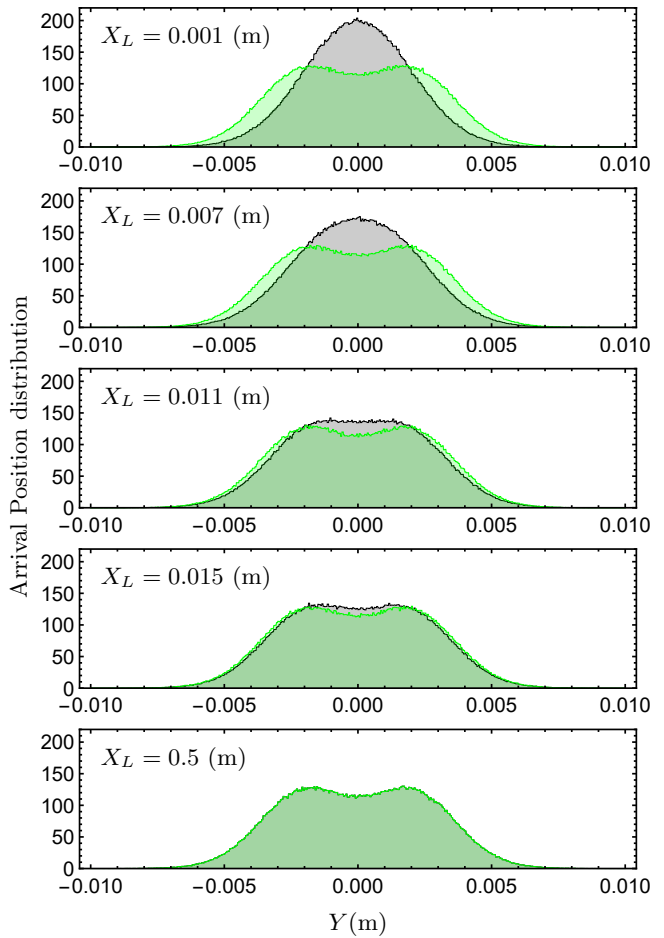\label{non-eq}
In Bohmian formalism, non-locality is manifest due to non-local particle dynamics \cite{Quantum physics without quantum philosophy}. 
Moreover, here as we mentioned above, detection of the left moving particle makes \textit{instantaneous} deviation in the trajectory of the right one. At the statistical level, this can affect the joint distribution in $(y_L,y_R)$-space, which are represented by dark scatter plots in figure \ref{Joints-Marginals}. Nevertheless, the right spatial marginal distribution, as a local observable quantity, turns out to be independent of whether there is any screen on the left or not and if there is any, it is not sensitive to the location of that detection screen. This is compatible with the well-known non-signaling theorem \cite{Ghirardi1980}. It is astonishing that not only the non-local Bohmian dynamics is not leading to faster than light signaling, but on the contrary, it is vital for producing such fine-tuned mixing of superpositions and therefore saving signal-locality.  
It should also be noted, that the usual non-signaling theorem is proved for observable quantities which are described by self-adjoint operators in the standard formalism, while the marginal position distribution here is not of this type \cite{foot5}. 
Nonetheless, our numerical study, based  on Bohmian mechanics, approve signal-locality for this non-standard, but experimentally observable, distributions \cite{foot6}.

%However, the important point here is that the marginal spatial distribution can not be computed in the standard quantum mechanics as it needs the arrival time data.
%The usual  non-signaling theorem is proved for local-distribution which given by POVM.

% Note that, in conventional quantum mechanics Born initial distribution is taken to be an axiom or law of nature. Whereas in pilot-wave theory it is a special state of ‘quantum equilibrium’ \cite{}. The Born probability distribution is assumed as an initial condition. Infact, the theory allows one to consider arbitrary initial distributions which violate quantum theory [5, 10, 13, 14].There are currently two main approaches to understanding the Born rule in pilot-wave theory. In the one of this approach, which developed by A. Valentini and collaborators, the Born rule we observe today is explained by a process of "dynamical relaxation", whereby initial nonequilibrium distributions evolve towards equilibrium on a coarsegrained level.

The signal-locality here is not however a result of the Bohmian dynamics alone and the Born rule for initial particle position distribution, i.e. quantum equilibrium condition, is critical as well \cite{Valentini1991,Valentini2002}.  To test this fact here, we make a deviation from Born rule by generating the initial positions through the same mixture of Gaussian distributions as $|\Psi_0(\bm R^0)|^2$ except with replacing $\sigma \rightarrow  \sigma/2$. The result is shown in figure \ref{Deviation from Born Interference-fig} where we have plotted the right screen marginal distributions for different locations of the left screen. First, note that with deviation from the Born rule, we should not expect that the right green marginal (even without considering the collapse effect) should be the same as before. But, the point we want to emphasize here is that how in contrast to the quantum equilibrium case the right true marginal (with considering collapse) is dependent on the location of left detection screen and so, deviation from Born rule leads to faster than light signaling. 

%Note that deviation from Born rule is theoretical here and possible only in the simulation as we do not have direct control over it in a real experiment. Nevertheless, this argument could be reversed and one can check Born rule indirectly by checking signal-locality in double-double slits setup. This reversion 

%This fact can be used as a new way to experimentally test of the Born rule in multi-slit experiment \cite{Pleinert2020}; In principle, an upper bound on the deviation of signal-locality could be related to an upper bound on the Born rule. 

\textit{Summary and outlook.---}In any generic quantum experiment that contains some entangled particles and some particle detectors waiting for them to arrive, there is an obstacle to computing the time evolution of the quantum state in orthodox formalism from start to the end, when all the particles will be detected. The main reason is the absence of a unique prediction of the middle detection times in the orthodox formalism. The double-double-slit experiment here is one of the simplest examples of this kind which we have analyzed using the Bohmian formalism that can predict the quantum state to the end without any ambiguity. The main result is the joint detection distribution in a generic setup which is in fact a new experimentally testable prediction beyond orthodox quantum mechanics.
%Moreover, we have observed and approved signal-locality for spatial marginal distribution for each parties as a non-standard quantum observable.
One should note, however, that Bohmian mechanics is not the only framework that can predict this quantity and actually any interpretation with a proposal to arrival time such as decoherent histories \cite{Anastopoulos2017,Anastopoulos2012,Halliwell2009}, Nelson stochastic mechanics \cite{Nitta2008} and other approaches \cite{Maccone2020,Das2021Questioning,Marchewka,Nikolic2020} can have predictions in principle. This fact opens a new window to compare different interpretations of quantum theory. In this regard, the study of double-double-slit experiment in these frameworks would be an interesting extension of the present work, which has been left for future studies.
%Moreover, the signal-locality as a theoretical constraint may rule out some of these proposals.  Here, while consistency of the Bohmian results with signal-locality is shown using a numeric simulation, an analytical proof is also demanded, which has been left for future studies. 
The other interesting improvement is to consider the back effect of presence of the detection screen on time evolution of the wave function. The arrival distributions computed here, should be considered as \textit{ideal} or \textit{intrinsic} distributions, since that effect is ignored \cite{Das2019Exotic}. The influence of a physical detector can be modeled in various phenomenological methods such as using a complex potential \cite{Allcock1969II,Halliwell2008}, absorbing boundary condition \cite{Tumulka2022} and other approaches \cite{Marchewka}. In principle, these methods can lead to different predictions where the comparison would be interesting and has been left for future works, as well.

\textit{Acknowledgment.---}The authors are grateful to Ali Ayat and Reinhard Werner for their useful discussions and comments and to Hamidreza Safari for his support of this work. The work of VH is supported in part by Iran Science Elites Federation under Grant No. 401138.


\begin{thebibliography}{9999999999999}
\bibitem{Horne1993}
M. A. Horne and A. Zeilinger, 
%Multiparticle Interferometry and the Superposition Principle, 
\href{https://doi.org/10.1063/1.881360}{Physics Today 46, 8, 22 (1993)}. 
\bibitem{Horne1989}
M. A. Horne,  A. Shimony,  and A. Zeilinger, 
%Two-particle interferometry, 
\href{https://link.aps.org/doi/10.1103/PhysRevLett.62.2209}{Phys. Rev. Lett, $\bm{62}$ (1989)}.
\bibitem{Braverman2013}
B. Braverman  and C. Simon, 
%Proposal to Observe the Nonlocality of Bohmian Trajectories with Entangled Photons, 
\href{https://link.aps.org/doi/10.1103/PhysRevLett.110.060406}{Phys. Rev. Lett. $\bm{110}$ (2013)}.
\bibitem{Kaur2020}
M. Kaur and M. Singh, 
%Quantum double-double-slit experiment with momentum entangled photons, 
\href{https://doi.org/10.1038/s41598-020-68181-1}{Sci Rep
$\bm{10}$ (2020)}.
\bibitem{Waitz2016}
M. Waitz, and et al, 
%Two-Particle Interference of Electron Pairs on a Molecular Level, 
\href{https://link.aps.org/doi/10.1103/PhysRevLett.117.083002}{Phys. Rev. Lett, $\bm{117}$ (2016)}.
\bibitem{Kofler2012}
J. Kofler and et al, 
%Einstein-Podolsky-Rosen correlations from colliding Bose-Einstein condensates, 
\href{https://link.aps.org/doi/10.1103/PhysRevA.86.032115}{Phys. Rev. A, $\bm{86}$ (2012)}.
\bibitem{Allcock1969I}
G. R. Allcock,
%The time of arrival in quantum mechanics I. Formal considerations,
\href{https://doi.org/10.1016/0003-4916(69)90251-6}{Annals of Physics $\bm{53}$, (1969)}.
\bibitem{Kijowski1974}
J. Kijowski, 
%On the time operator in quantum mechanics and the Heisenberg uncertainty relation for energy and time, 
\href{https://doi.org/10.1016/S0034-4877(74)80004-2}{Reports on Mathematical Physics $\bm{6.3}$, (1974)}.
\bibitem{Werner1985}
R. Werner,  
%Screen observables in relativistic and nonrelativistic quantum mechanics, 
\href{https://doi.org/10.1063/1.527184}{Journal of mathematical physics $\bm{27.3}$ (1986)}.
\bibitem{Vona2013}
N. Vona, G. Hinrichs, and D. D\"{u}rr, 
%What does one measure when one measures the arrival time of a quantum particle?, 
\href{https://journals.aps.org/prl/abstract/10.1103/PhysRevLett.111.220404}{Phys. Rev. Lett, $\bm{111}$,  (2013)}.
\bibitem{Das2021gauge}
Das, Siddhant, and Markus Noth, 
%Times of arrival and gauge invariance, 
\href{https://doi.org/10.1098/rspa.2021.0101}{Proceedings of the Royal Society A, $\bm{477}$ (2021)}.
\bibitem{Das2021Questioning}
S. Das and W. Struyve, 
%Questioning the adequacy of certain quantum arrival-time distributions, 
\href{https://link.aps.org/doi/10.1103/PhysRevA.104.042214}{Phys. Rev. A, $\bm{104}$ (2021)}.
\bibitem{Maccone2020}
L. Maccone,  and K. Sacha,  
%Quantum Measurements of Time, 
\href{https://link.aps.org/doi/10.1103/PhysRevLett.124.110402}{Phys. Rev. Lett, $\bm{124}$ (2020)}.
\bibitem{Anastopoulos2017}
C. Anastopoulos,  and N. Savvidou,  
%Time-of-arrival correlations, 
\href{https://link.aps.org/doi/10.1103/PhysRevA.95.032105}{Phys. Rev. A,  $\bm{95}$ (2017)}.
%%%%%%%%%%%%%%
\bibitem{footPauli}
As shown by Pauli \cite{Pauli}, if the Hamiltonian spectrum is discrete or has a lower bound, then a self-adjoint time
operator, canonically conjugate to the Hamiltonian, does not exist.
\bibitem{Pauli}
 W. Pauli, in: S. Flugge (Ed.), Encyclopedia of Physics, vol. 5/1, Springer, Berlin, p. 60,
(1958).
%%%%%%%%%%%coincidennt double double slit
\bibitem{Georgiev2021}
D. Georgiev,  and et al, 
%One-particle and two-particle visibilities in bipartite entangled Gaussian states, 
\href{https://link.aps.org/doi/10.1103/PhysRevA.103.062211}{Phys. Rev. A, $\bm{103}$ (2021)}.
\bibitem{Gneiting2013}
C. Gneiting,  and K. Hornberger, 
%Nonlocal Young tests with Einstein-Podolsky-Rosen-correlated particle pairs, 
\href{https://link.aps.org/doi/10.1103/PhysRevA.88.013610}{Phys. Rev. A, $\bm{88}$ (2013)}.
\bibitem{Guay12003}
E. Guay1 and L. Marchildon, 
%Two-particle interference in standard and Bohmian quantum mechanics, 
\href{https://doi.org/10.1088/0305-4470/36/20/317}{J. Phys. A: Math. Gen. $\bm{36}$ (2003)}.
\bibitem{Fonseca2000}
E. J. S. Fonseca,  and t al. 
%Controlling two-particle conditional interference, 
\href{https://link.aps.org/doi/10.1103/PhysRevA.61.023801}{Phys. Rev. A, $\bm{61}$ (2000)}.
%%%%%%%%%%%%%%%%%%%%%%%% Bohmian-standard
\bibitem{Bohm1952}
D. Bohm, 
%A suggested interpretation of the quantum theory in terms of "hidden" variables. I \& II, 
\href{https://link.aps.org/doi/10.1103/PhysRev.85.166}{Phys. Rev. $\bm{85}$ (1952)}; \href{https://link.aps.org/doi/10.1103/PhysRev.85.180}{Phys. Rev. $\bm{85}$ (1952)}.
\bibitem{Durr2004}
D. D\"{u}rr, S. Goldstein, and N. Zanghi, 
%Quantum equilibrium and the role of operators as observables in quantum theory, 
\href{https://doi.org/10.1023/B:JOSS.0000037234.80916.d0}{Journal of Statistical Physics $\bm{116}$ (2004)}.
%%%%%%%%%%%%%%%%%%%%%%%% in so far
\bibitem{Bell1995}
J. S. Bell, 
%de broglie-bohm, delayed-choice, double-slit experiment, and density matrix,
% in Quantum Mechanics, High Energy Physics And Accelerators: 
\href{https://doi.org/10.1142/9789812795854_0083}{Selected Papers Of John S Bell, World Scientific,  pp. 788-792, (1995)}.
\bibitem{Ivanov2017}
I. Ivanov, C. H. Nam, and K. T. Kim, 
%Exit point in the strong field ionization process, 
\href{https://doi.org/10.1038/srep39919}{Sci Rep $\bm{7}$ (2017)}.
\bibitem{Das2019}
S. Das and D. D\"{u}rr, 
%Arrival time distributions of spin-1/2 particles, 
\href{https://doi.org/10.1038/s41598-018-38261-4}{Sci Rep $\bm{9}$ (2019)}.
%%%%%%%%%%%%%%%%%%%%%%%% Time in Bohmian
\bibitem{Leavens1998}
C. R. Leavens, 
%Time of arrival in quantum and Bohmian mechanics, 
\href{https://link.aps.org/doi/10.1103/PhysRevA.58.840}{Phys. Rev. A, $\bm{58}$ (1998)}.
\bibitem{Zimmermann2016}
T. Zimmermann and et al, 
%Tunneling time and weak measurement in strong field ionization, 
\href{https://link.aps.org/doi/10.1103/PhysRevLett.116.233603}{Phys. Rev. lett, $\bm{116}$ (2016)}.
%%%%%%%%%%%%
\bibitem{Durr2010On}
D. D\"{u}rr, et al. 
%On the Quantum Mechanical Scattering Statistics of Many Particles, 
\href{https://doi.org/10.1007/s11005-010-0404-6}{Lett Math Phys, $\bm{93}$  (2010)}.
\bibitem{Quantum physics without quantum philosophy}
D. D\"{u}rr and et al. 
%Quantum physics without quantum philosophy. 
\href{https://link.springer.com/book/10.1007/978-3-642-30690-7}{Springer Science \& Business Media, (2012)}. 
\bibitem{Norsen2014} 
T. Norsen, and W. Struyve, 
%Weak measurement and Bohmian conditional wave functions, 
\href{https://doi.org/10.1016/j.aop.2014.07.014}{Annals of Physics $\bm{350}$ (2014)}.
%%%%%%%%%%%%%%%%%%%%%%%%%%%%%%%%%%%%%%%%
\bibitem{Perrin2007}
A. Perrin and et al. 
%Observation of Atom Pairs in Spontaneous Four-Wave Mixing of Two Colliding Bose-Einstein Condensates, 
\href{https://link.aps.org/doi/10.1103/PhysRevLett.99.150405}{Phys. Rev. lett, $\bm{99}$ (2007)}.
%%%%%%
\bibitem{Kocsis2011} 
S. Kocsis and et al, \href{https://www.science.org/doi/10.1126/science.1202218}{Science $\bm{332}$ (2011)}.
\bibitem{Mahler2014} 
D. H. Mahler, et al. 
%Measuring Bohm trajectories of entangled photons, 
\href{https://doi.org/10.1364/CLEO_QELS.2014.FW1A.1}{CLEO: QELS Fundamental Science. Optica Publishing Group, (2014)}.
\bibitem{Mahler2016} 
D. H. Mahler, et al. 
%Experimental nonlocal and surreal Bohmian trajectories, 
\href{https://www.science.org/doi/full/10.1126/sciadv.1501466}{Science advances  $\bm{2.2}$ (2016)}.
\bibitem{Xiao2017} 
Xiao, Ya, et al. 
%Experimental nonlocal steering of Bohmian trajectories, 
\href{https://doi.org/10.1364/OE.25.014463}{Optics Express 25.13 (2017)}.
\bibitem{Traversa2013} 
F. L. Traversa,  and et al.
%Robust weak-measurement protocol for Bohmian velocities,
\href{https://link.aps.org/doi/10.1103/PhysRevA.87.052124}{Phys. Rev. A, $\bm{87}$ (2013)}.
%%%%%%%%%%
\bibitem{foot5}
Note that, the Heisenberg position operator describe position measurement at a specific time, not position measurements at random times \cite{Durr2004,DurrTeufel2004}.
\bibitem{DurrTeufel2004} 
D. D\"{u}rr,  and S. Teufel, 
%On the Exit Statistics Theorem of Many-particle Quantum Scattering, 
\href{https://doi.org/10.1007/978-0-8176-8202-6_4}{Multiscale Methods in Quantum Mechanics. Trends in Mathematics. (2004)}. 
\bibitem{Ghirardi1980} 
 G. C. Ghirardi and et al., 
 %general argument against superluminal transmission through the quantum mechanical measurement process, 
 \href{ https://doi.org/10.1007/BF02817189}{Lett. Nuovo Cimento 27, 293–298 (1980)}.
 %%%%%%
\bibitem{foot6} 
Here, while consistency of the Bohmian results with signal-locality is shown using a numeric simulation, an analytical proof is also demanded, which has been left for future studies. 
%%%%%%%%%%%%%
\bibitem{Valentini1991} 
A. Valentini,  
%Signal-locality, uncertainty, and the subquantum H-theorem. II, 
\href{https://doi.org/10.1016/0375-9601(91)90330-B}{Physics Letters A $\bm{158}$ (1991)}.
\bibitem{Valentini2002} 
A. Valentini, 
%Signal-locality in hidden-variables theories, 
\href{https://doi.org/10.1016/S0375-9601(02)00438-3}{Physics Letters A $\bm{297}$ (2002)}.

%%%%%
\bibitem{Anastopoulos2012}
C. Anastopoulos and N.Savvidou, 
%Time-of-arrival probabilities for general particle detectors, 
\href{https://link.aps.org/doi/10.1103/PhysRevA.86.012111}{Phys. Rev. A, $\bm{86}$ (2012)}.
\bibitem{Halliwell2009}
J. J. Halliwell and J. M. Yearsley, 
%Quantum arrival time formula from decoherent histories, 
\href{https://doi.org/10.1016/j.physleta.2009.10.077}{Phys. Lett. A, $\bm{374}$ (2009)}.
%%%%%
\bibitem{Nitta2008}
H. Nitta and T. Kudo, 
%Time of arrival of electrons in the double-slit experiment, 
\href{https://link.aps.org/doi/10.1103/PhysRevA.77.014102}{Phys. Rev. A, $\bm{77}$ (2008)}.
\bibitem{Nikolic2020}
D. Jurman and H. Nikolic, 
%The time distribution of quantum events, 
\href{https://doi.org/10.1016/j.physleta.2021.127247}{Phys. Lett. A,  $\bm{396}$ (2020)}.
\bibitem{Marchewka}
A. Marchewka, and Z. Schuss, 
%Feynman integrals with absorbing boundaries, 
\href{https://doi.org/10.1016/S0375-9601(98)00107-8}{Phys. Lett. A, $\bm{240}$ (1998)};
%Measurement as absorption of Feynman trajectories: Collapse of the wave function can be avoided, 
\href{https://link.aps.org/doi/10.1103/PhysRevA.65.042112}{Phys. Rev. A  $\bm{65}$ (2002)}.

%%%%%%%
\bibitem{Das2019Exotic}
S. Das, M. N\"{o}th, and D. D\"{u}rr, 
%Exotic Bohmian arrival times of spin-1/2 particles: an analytical treatment,
\href{https://link.aps.org/doi/10.1103/PhysRevA.99.052124}{Phys. Rev. A $\bm{99}$ (2019)}.
\bibitem{Allcock1969II}
G. R. Allcock, 
%The time of arrival in quantum mechanics II. The individual measurement, 
\href{https://doi.org/10.1016/0003-4916(69)90252-8}{Ann. Phys. (NY) $\bm{53}$ (1969)}.
\bibitem{Halliwell2008}
J. J. Halliwell, 
%Path-integral analysis of arrival times with a complex potential, 
\href{https://link.aps.org/doi/10.1103/PhysRevA.77.062103}{Phys. Rev. A $\bm{77}$ (2008)}.
\bibitem{Tumulka2022}
R. Tumulka,
%Distribution of the time at which an ideal detector clicks, 
\href{https://doi.org/10.1016/j.aop.2022.168910}{Annals of Physics,  $\bm{442}$ (2022)}.
\end{thebibliography}
\end{document}